\documentclass[twocolumn]{elsart}
\begin{document}
\begin{frontmatter}  


\title{The MECO experiment: A search for \\
lepton flavor violation in muonic atoms}
\author{James L. Popp for the MECO Collaboration}

\address{Department of Physics, New York University,
New York, NY 10003}
\date{\today}
\maketitle

%
%
\begin{abstract}
MECO will search for direct evidence of muon and electron
flavor violation in the decay of muons in Coulomb bound states
via coherent recoil of the nucleus and decay electron.
The expected sensitivity to the
$\mu^{-} {\rm N} \rightarrow e^{-} + {\rm N}$
branching fraction relative to muon capture
$\mu^{-} {\rm N}(A,Z) \rightarrow \nu_{\mu} + {\rm N}(A,Z-1)$ is
$R_{\mu e} < 5\times 10^{-17}$ at $90\%$ confidence level, roughly
three to four orders of magnitude lower than current limits.  This
article provides an overview of the experiment.
\end{abstract}
\end{frontmatter}  

%
%
\section{Introduction}
Lepton-flavor-violating transition searches are likely to be
sensitive probes of super-unified theories since the predicted
rates are often nearest experimental limits.  Although precise
predictions of theories depend on the specific model, the physical
mechanisms that lead to lepton flavor violations are generic to 
supersymmetric quark-lepton unification.
Many extensions to the Standard Model that unify quarks and leptons,
including supersymmetric theories, e.g., \cite{ref:Barbieri_Hall},
suggest that in muonic atoms the branching fraction for coherent
conversion of a muon into an electron relative to muon capture in
the nucleus $R_{\mu e} \sim 10^{-14} - 10^{-17}$ over much of
the parameter space.  In terms of decay rates, with mass number
$A$ and atomic number $Z$, 
$R_{\mu e} =
\Gamma (\mu^{-} {\rm N}(A,Z) \rightarrow e^{-}     + {\rm N}(A,Z))/
\Gamma (\mu^{-} {\rm N}(A,Z) \rightarrow \nu_{\mu} + {\rm N}(A,Z-1))$,
where the numerator only involves the internal nuclear ground state
and denominator includes all possible nuclear final states.
The MECO experiment~\cite{ref:MECO_collaboration} will
search for direct evidence for this lepton-flavor-violating process.

The signature of coherent muon-electron conversion is a two-body
final state with a mono-energetic $E_{\rm o}$ electron.
When the internal state of the nucleus remains in the ground state
during the transition the electron recoils coherently off the entire nucleus,
resulting in a strong enhancement in the conversion rate.
The electron energy
$E_{\rm o} \simeq E_{\mu} + E_{\mu}^2/2M_{\rm N}$ with
$E_{\mu} = m_{\mu} + {\rm BE}$, the muon mass plus Coulomb binding energy
${\rm BE}$. 
The Schr{\"{o}}dinger equation gives good estimates of ${\rm BE}$
for light nuclei; neglecting the size of the nucleus in aluminum,
the MECO stopping target material, ${\rm BE(1s)} \sim -0.48\,$MeV and
$E_{\rm o} \sim 105.0\,$MeV.
MECO will look for conversion electrons in the energy window
$103.6$-$105.1\,$MeV, where the signal to background ratio is
$\sim 20$.

The dominant background as in previous
experiments~\cite{ref:Sindrum_II,ref:TRIUMF} is muon decay in orbit
$\mu^{-} {\rm N} \rightarrow \nu_{\mu} + {\bar {\nu}}_{e} + e^{-} + N$;
the electron energy spectrum falls rapidly near the endpoint
at exactly the conversion energy, $\propto (E_{\rm o} - E_{e})^5$.
Thus background rejection improves rapidly with energy resolution.
The MECO tracker probes this crucial region more deeply than ever
before. 

The current measured limit set by Sindrum II~\cite{ref:Sindrum_II}
is $R_{\mu e} < 6.1\times 10^{-13}$ at $90\%$ confidence in titanium,
with sensitivity expected to improve soon to $2\times 10^{-14}$.
This experiment is limited by beam intensity, not background.
With a significantly scaled-up approach to muon production and
transport, $\sim 1.8\times 10^{11}\,\mu^{-}/$s at the stopping
target and at least $10^4$ greater than existing low-energy
($p < 100\,{\rm MeV}/{\rm c}$) muon beams, MECO is expected to
improve on these limits by at least three to four orders of magnitude
$R_{\mu e} < 5\times 10^{-17}$.

%
%
\section{The experiment}
\begin{figure}[tb!]
\begin{center}
\vspace{4.0 cm} 
\includegraphics{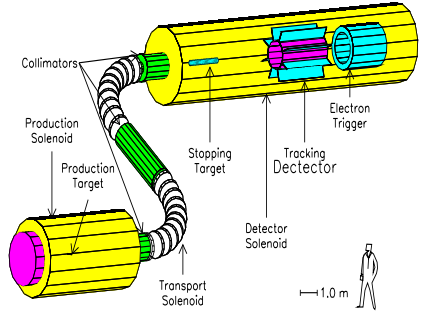}
\end{center}
\caption{The MECO experiment}
\label{figure:MECO_setup}
\end{figure}

The MECO setup is shown in Figure~\ref{figure:MECO_setup} and consists of
three large superconducting magnets: the
muon production, transport, and detector solenoids. 
The clear bores of the magnets are connected forming a single
cavity, held in vacuum.
The magnetic field is continuous throughout the cavity.
In the $4\,$m long production solenoid, the
axial magnetic field intensity $B$ is graded
from $3.6\,$T to $2.3\,$T.  The field follows the $12\,$m S-shaped
transport solenoid, falling only in the straight sections
from $2.3\,$T to $2.0\,$T.
In the $10.5\,$m detector solenoid $B$ is graded from $2.0\,$T at the 
entrance to $1.0\,$T between the stopping target and tracking
detector, and constant thereafter.
Both the trigger and tracker are high-rate detectors with their
detection material positioned far from the solenoid axis to intercept
conversion electron helices and avoid beam particle interactions.
%
%

The primary proton beam is provided by the Alternating
Gradient Synchrotron (AGS), with $\sim 1\,$sec machine cycle for
$40\times 10^{12}$ protons.  The choice of beam energy,
$7$-$8\,$GeV, optimizes $\pi$ and minimizes $\bar p$
(a potential background source) production.
The beam (not shown) enters the production solenoid on
the right at $10^{\rm o}$ to the axis and exits to the left.
A tungsten target is bombarded to produce $\pi^{\pm}$ which
subsequently decay to $\mu^{\pm}$, 
following helical trajectories along the magnet axis.
The graded field reflects left-moving charged particles with
momenta outside the $30^{\rm o}$ (half angle) loss cone,
toward the transport entrance.
Since particle energy and $p_{\perp}^2 / B$ are constant,
as charges move to regions of lower $B$
the angle $\chi$ between the momentum and solenoid axis decreases.
Pions with transverse momenta $p_{\perp} < 180\,{\rm MeV}/{\rm c}$ travel
within the $30\,$cm inner radius of the magnet.  Most $\pi$ decays occur
in the production region.

A pulsed beam with good extinction $\varepsilon$
between pulses is critical to prompt background rejection and 
thus success for MECO.
The mean lifetime for $\mu^{-}$ in aluminum atoms
$\tau_{\mu}^{\rm Al} = 880\,$ns sets the pulse separation time scale.
Thus the pulse spacing is microseconds and given by
$\tau_{\rm pulse} + \tau_{\rm delay} + \tau_{\rm obs}$.
The actual proton pulse time $\tau_{\rm pulse} \ll \tau_{\mu}^{\rm Al}$.
Prompt backgrounds are reduced to an acceptable level by, first,
introducting a delay $\tau_{\rm delay}$ between each proton pulse
and the detection window $\tau_{\rm obs}$ so that all beam particles
have left the detector region before $\tau_{\rm obs}$; and second,
by requiring that the ratio of protons crossing the production target
during the observation window to that during the pulse be
$\varepsilon \sim 10^{-9}$.  AGS extinction studies are ongoing with
current measured value $\varepsilon \sim 10^{-7}$.

The transport solenoid filters the particle flux producing a
momentum- and charge-selected muon beam, with good reduction in
contamination from $e^{\pm}$, $\mu^{+}$, $\pi^{\pm}$, $p$, and $\bar p$.
Transport entrance and exit collimators with $15\,$cm inner radius further
limit the beam.  For sufficiently low momenta, centers of particle orbits
follow the S-shape of the solenoid;
however, the curvature of the magnetic field in the two bent sections  
gives rise to a drift out of the plane of the magnet.
In the first section positives drift up and negatives down, and
in the second directions are reversed.  
By vertically limiting the aperture of a narrower 
collimator in the center straight section with a downward offset 
positives and high-energy negatives are absorbed from the beam.
Low-energy antiprotons originating in the production region
have long transport crossing times, and as already mentioned
the potential to create backgrounds at the detectors.
A thin beryllium window in the center section reduces the $\bar p$
component to a negligable level.  The shallow grades in $B$ in each
straight section further prevents late arrivals.  At the transport exit
$e^{+}$ and $\mu^{+}$ have $p < 40\,{\rm MeV}/{\rm c}$, most $e^{-}$
have $p < 80\,{\rm MeV}/{\rm c}$ with none above $100\,{\rm MeV}/{\rm c}$.
The $\mu^{-}$ beam has $14\,{\rm MeV}/{\rm c} < p < 100\,{\rm MeV}/{\rm c}$. 

Only $\mu^{-}$ below $50\,{\rm MeV}/{\rm c}$ stop in the target.
Negative muons come to rest in matter by slowing to thermal-like
velocities through inelastic atomic collisions and falling into a
Coulomb orbit about a nucleus.  Excited states cascade to lower
ones with time scale $10^{-13}\,$s producing X-rays and Auger electrons.
Approximately $60\%$ are captured by the nucleus and
$40\%$ decay in orbit.  

The stopping target design optimizes the probability for
muon stopping and conversion electron detection, and
minimizes energy loss of exiting signal $e^{-}$ and the number of
decay-in-orbit electrons reaching the detectors.  Background rates
are reduced by minimizing stopping target mass.
The target has $17$ parallel disks $0.02\,$cm thick with
$5.0\,$cm spacing and radii from $8.3$-$6.5\,$cm; total mass is $159\,$g.
Simulations yield a stopping efficiency $0.0025\,\mu^{-}$ per proton,
i.e., $10^{11}$Hz stop rate.

The target is centered in the graded field between
$1.63$-$1.37\,$T; the gradient reflects $e^{-}$ emitted upstream
back to the detectors, resulting in $\sim 60\%$ of all conversions
hitting the tracker.  
Conversion electrons with $p_{\perp} > 90\,{\rm MeV}/{\rm c}$
($120^{\rm o} > \chi > 60^{\rm o}$) 
are swept forward by the decreasing magnetic field into the range
$75\,{\rm MeV}/{\rm c} < p_{\perp} < 90\,{\rm MeV}/{\rm c}$
($45^{\rm o} < \chi < 60^{\rm o}$) at the tracking detector.
Beam particles with $p_{\perp} < 90\,{\rm MeV}/{\rm c}$ that do not
scatter in the target pass down the center of the 
solenoid without intercepting the detectors.
Placing the target in a graded field also ensures that $e^{-}$
originating upstream of the gradient with
$105\,{\rm MeV}/{\rm c}$ arrives at the detectors with
$p_{\perp} < 75\,{\rm MeV}/{\rm c}$, eliminating many potential
backgrounds.
On average the stopping target produces $2\,n$, $2\,\gamma$,
and $0.1\,p$ per muon capture, displacing the detectors $>1\,$m
downstream from the target greatly reduces acceptance for
$n$ and $\gamma$. 

The straw tube drift chamber tracking detector has large acceptance
for and measures with good efficiency the helix parameters of 
conversion electrons in the uniform field.  
Good energy resolution is essential to
distinguish conversion electrons from decay in orbit,
$\raisebox{-0.5ex}{$\stackrel{<}{\sim}$} 900\,$keV FWHM. 
Resolution is dominated by scattering in the tracker,
and to a much lesser extent by pattern recognition errors. 
Energy loss in the target, proton absorbers (not shown), 
and tracker broadens the peak of the resolution function
for conversions, and introduces a small mean energy loss and
low-energy tail. 

The tracker has 16 rectangular planes, 8 forming an octagon
centered on the solenoid axis and 8 positioned at each corner
extending radially.  This geometry gives $\ge 3$ plane crossings
per conversion $e^{-}$ helix orbit; $3\,$m long detector planes
guarantee $\ge 2$ helix orbits in the detector, a powerful advantage for
pattern recognition.  With tubes parallel to the length of a plane,
a slight tilt of each plane, $< 1^{\rm o}$, prevents a conversion
$e^{-}$ from entering the same tube on different orbits.

Each plane has three layers of $0.25\,$cm radius circular tubes
in a close-packed arrangment; each tube contains a central anode wire.
Drift time information and wire positions combine to give the
track-detector plane crossing
position perpendicular to the wires and the track projection angle
(valuable for pattern recognition) in a plane perpendicular to the wires
and wire planes.  Position along the wires is measured with
capacitively-coupled cathode foils, $\sim 1\,$cm wide,
running perpendicular to the wires on both sides of a plane.
Resolutions from wire information are highly angle-dependent
with average position and angle resolutions, expressed as root
mean square (RMS) quantities, $0.014\,$cm and $1.0^{\rm o}$. 
The expected RMS pad resolution is $\sim 0.05\,$cm.

The purpose of the trigger calor\-imeter is to select
with high efficiency conversion electrons
that pass through the tracker,
while minimizing triggers from lower-energy events.
A high degree of segmentation limits signal pile-up.
Since background rises exponentially as the trigger threshold decreases, 
it is important to have $E_{\rm threshold}$ as high as possible.
The calorimeter should also have good energy and position resolution to
provide an independent measurement of these quantities for the helix,
a feature shown to have high discriminant value to suppress background.

Trigger design is currently under development.  The current design
(not shown) using single 
$3\times 3\times 12\,{\rm cm}^3$ crystals of BGO ($300\,$ns
EM shower decay constant) and avalanche photodiodes
show that $E_{\rm threshold} = 80\,$MeV gives $87\%$ efficiency for
conversions.  This results in an acceptable $0.2\,$kHz trigger rate.
The RMS position and energy resolutions for $105.0\,$MeV electrons
are $1.0\,$cm and $5.3\,$MeV.   Research continues on other materials
with shorter decay times, such as GSO and ${\rm PbWO}_4$.

A conversion electron candidate event has: (1) detector
signals within the observation time window, (2)
calorimeter shower energy $\ge E_{\rm threshold}$, (3) detector plane
crossings found by pattern recognition consistent with a helix
of $\ge 2$ helix orbits, (4) a helix which when extended back
intersects the stopping target and extended forward is consistent
with the calorimeter shower position, (5) $p_{\perp}$ of the helix
in the range $75\,{\rm MeV}/{\rm c} < p_{\perp} < 90\,{\rm MeV}/{\rm c}$,
(6) helix energy consistent with calorimeter shower energy, and
(7) helix energy between $103.6$-$105.1\,$MeV.

%
%
\section{Conclusion}

Table~\ref{table:meco_backgrounds}
\begin{table}[tb!]
\begin{center}
\begin{tabular}{ll}
\hline
Source                               & Events       \\
\hline
$\mu^{-}$ DIO                        & $  0.25$     \\
Pattern Recognition                  & $< 0.006$    \\ 
$\mu^{-}$ RC                         & $< 0.005$    \\
$\star$ $\mu^{-}$ DIF (a)            & $< 0.03$     \\
$\star$ $\mu^{-}$ DIF (b)            & $  0.04$     \\
$\star$ $\pi^{-}$ RC\phantom{P}      & $  0.07$     \\
$\star$ $\pi^{-}$ DIF                & $< 0.001$    \\
$\star$ beam $e^{-}$                 & $< 0.04$     \\
$\pi^{-}$ RC\phantom{P}              & $  0.001$    \\ 
$\bar{p}$ induced                    & $  0.007$    \\
CR induced                           & $  0.004$    \\
\hline
Total                                & $< 0.45$     \\
\hline
\end{tabular}
\end{center}
\vspace{0.1in}
\protect
\caption{Expected MECO backgrounds (abbreviations in text)
for $\varepsilon = 10^{-9}$ and $10^{7}$s running time.
Stars indicate extinction dependence.} 
\label{table:meco_backgrounds}
\end{table}
summarizes calculations of primary background contributions
in the energy window $103.6$-$105.1\,$MeV for a running time of
one-third of a year.  Decay in orbit (DIO) and radiative 
capture (RC) of $\mu^{-}$ are intrinsic to muon decay,
both are made negligible with precise energy measurement. 
Starred entries arise from out-of-time proton crossings at
the production target which are proportional to the extinction:
Decay in flight (DIF) of $\mu^{-}$ without (a) and with (b) 
scattering in the stopping target, radiative pion capture,
DIF for $\pi^{-}$, and beam $e^{-}$.
The next two backgrounds in the table are from source particles with
long transit times in the transport solenoid.  Acceptable
cosmic ray (CR) induced background rejection for MECO requires
passive and active shielding with a modest improvement
over previous~\cite{ref:Sindrum_II,ref:TRIUMF}.
Inefficiency in the CR shield veto is expected to be $10^{-4}$. 
Assuming $\varepsilon \sim 10^{-9}$
the total number of background events expected is $< 0.45$.

Table~\ref{table:meco_sensitivity}
shows that a four-month run and $R_{\mu e} = 10^{-16}$ 
yields $\sim 5$ muon-electron conversion events.
The number of muon captures is just the product of
the first seven entries of Table~\ref{table:meco_sensitivity},
where $P(\mu^{-}\,{\rm capture})$ is the probability of $\mu^{-}$ capture,
$F(\tau_{\rm obs},\,\mu^{-} {\rm capture})$
the fraction of $\mu^{-}$ capture in the observation window,
and the overall fitting and event selection criteria efficiency
is given.
%
%
\begin{table}[tb!]
\begin{center}
\begin{tabular}{ll}
\hline
Running time (s)                                   &  $10^7$             \\
Protons$/$s                                        &  $4 \times 10^{13}$ \\
\hline
$\mu^{-}$ stopped $/p$                             &  $0.0025$           \\
$P(\mu^{-}\,{\rm capture})$                        &  $0.60$             \\
$F(\tau_{\rm obs},\,\mu^{-} {\rm capture})$        &  $0.49$             \\ 
Trigger efficiency                                 &  $0.90$             \\
Fitting/selection                                  &  $0.19$             \\
$R_{\mu e}$                                        &  $10^{-16}$         \\
\hline
Detected events                                    &  $5$                \\
\hline
\end{tabular}
\end{center}
\vspace{0.1in}
\protect
\caption{Expected MECO sensitivity.} 
\label{table:meco_sensitivity}
\end{table}

%
%

\end{document}